\documentstyle[epsf,12pt]{article}
\begin{document}

\begin{center}
{\large {\bf Nonlinear fluctuation phenomena in the 
transport properties of superconductors}}

\medskip\medskip

{\bf A.I. Larkin$^{2,1}$ and Yu.N. Ovchinnikov$^1$}

\medskip\medskip

{\small {\it (1)\,\,\, L.D. Landau Institute for Theoretical Physics Academy
of Sciences of Russia, Kossigin Str. 2, 117940, Moskow, Russia }}
\end{center}

{\small {\it \bigskip }}

{\small {\it (2)\,\,\, Theoretical Physical Institute, University of
Minnesota, 116 Church Street SE, Minneapolis, Minnesota USA }}

\vspace{1cm}

{\small There  exist a wide temperature region $({\rm Gi}T<T-T_c<T\sqrt{{\rm Gi}}%
)$, where the influence of fluctuations on the thermodynamic properties of
superconductors can be taken into account in the linear (Gaussian)
approximation, while their influence on the kinetic properties is strongly
nonlinear. Maki-Thompson cotribution to conductivity saturates in this
region. However, Aslamazov-Larkin contribution becomes even more singular.
Such an enhancement is connected with the fact that nonlinear effects yield
an increase in the lifetime of fluctuating pairs. Pairbreaking and energy
relaxation processes can diminish the nonlinear effects. 
}

\section*{1. Introduction}

Electron scattering off the usual impurities leads to a temperature
independent residual resistance of the normal metal [1]. The conductivity of
bulk samples and films can be measured with a very high accuracy. This gives
a possibility to study different mechanisms, leading to the temperature
dependent conductivity at low temperatures. One of these mechanisms is
connected with the thermal fluctuations above the superconducting transition
temperature $T_c$ [2-5]. There are two kinds of fluctuation corrections,
leading to the temperature dependent conductivity above $T_c$. The first one
is known as Maki-Thompson contributions and the second one is conductivity of
fluctuating pairs (Aslamazov-Larkin contribution). These corrections have
different dependence on the spin flip scattering time $\tau_s$.
Characteristic temperature range for contributions of both types is Ginzburg
parameter ${\rm Gi}$, where ${\rm Gi}$ depends on dimensionality and for
films is equal to ${\rm Gi}=\tau_0= {1 / 32 \nu D d} =
e^2/16\hbar\sigma_{\Box}$. Where $\nu = m p^2 / 2 \pi^2$ is the electron
density of states per spin, $D={\frac{1 }{3}} v_F l_{tr}$ is the diffusion
coefficient, $d$ is the film thickness, $l_{tr}$ is the electron mean free
path, $p$ is the Fermi momentum, $\sigma_{\Box}$ is the conductance of a
square film. It has been found in paper [6], that nonlinear fluctuation
phenomena lead to a new temperature scale $T_c\sqrt{{\rm Gi}}$ (see also
[7-10]). In this paper we obtain expressions for conductivity in the
temperature region ${\rm Gi} <\tau <\sqrt{{\rm Gi}}$, where both the
Gaussian approximation works well and the nonlinear fluctuation effects are
important.

In paper [6] an attempt to find the fluctuating correction to conductivity
was made. The main point was that long wave fluctuations with $Dk^{2}<T \tau $
are essential. Such fluctuations can be considered as a Bose condensate.
Dynamics of superconductors should be considered on the background of these
fluctuations. They give a pseudogap in the excitation spectrum. In this
paper we will show, that short wave fluctuations with $Dk^{2}\gg T \tau $ can
be important. It was found in paper [11] that short wave fluctuations of the
order parameter $\Delta $ act on the electron Green functions as
paramagnetic impurities with depairing factor $\Gamma =\tau
_{s}^{-1}=\langle |\Delta |^{2}\rangle /\varepsilon $. Essential values of
energy $\varepsilon $ are of the order of $\varepsilon \sim \Delta \sim T%
\sqrt{\mbox {\rm {Gi}}}$. Hence $\Gamma $ is of the order of $T\sqrt{{\rm Gi}%
}$. So large value of the depairing factor leads to saturation of the M.-T.
contribution to conductivity in the temperature region $\tau <\sqrt{{\rm Gi}}$.

A more complicated situation takes place for A.-L. contribution. This
contribution is proportional to the density of pairs and their lifetime. For
large enough values of $\tau $ the time dependent Ginzburg-Landau equation
(TDGL) can be used to obtain this lifetime. It is proportional to $\hbar
/(T-T_{c})$ and hence A.L. contribution proportional to $\tau ^{-1}$. If the
concentration of paramagnetic impurities is large or if the energy
relaxation time is short, the TDGL equation can be used for all temperatures 
$T$. In this case A.L. contribution is valid in temperature range $\tau >{\rm %
Gi}$. But in the opposite limiting case, the nonlinear fluctuating effects
destroy the applicability of TDGL equation and lead to the increase of
lifetime of fluctuating pairs. As a result the A.L. contribution to
conductivity becomes more singular in the temperature region $\sqrt{{\rm Gi}}%
>\tau >{\rm Gi}$.

\section*{Qualitative picture}

In the temperature region $1\gg \tau \gg {\rm Gi}$ thermodynamic
fluctuations of the order parameter $\Delta $ can be considered to be
Gaussian. The
corresponding correlator has the form: 
\begin{equation}  \label{1}
\langle \Delta^*_k \Delta_k \rangle = \frac{T}{\nu d}~ 
\frac{1}{\tau + \frac{\pi D }{8T} k^2} = 
\frac{256}{\pi} \frac{{\rm Gi} T^2}{k^2 + 
\frac{8T\tau}{\pi D}}.
\end{equation}

In order to calculate thermodynamic quantities in the temperature region 
$\tau >{\rm Gi}$ it is sufficient to know this correlator only. However, to
calculate kinetic coefficients a more complicated problem has to be solved.
One has to find how Gaussian fluctuations change the one-particle excitation
spectrum. The longwavelength fluctuations with $k^{2}<k_{{\rm min}%
}^{2}=8T\tau /\pi D$ can be considered as a local condensate. They lead to
the formation of the pseudogap in the one-particle spectrum of excitations.
From Eq.(\ref{1}) we see that the pseudogap is equal to 
\begin{equation}
\Delta _{PG}={\frac{8}{\pi }}\sqrt{{\rm Gi}}T.  \label{2}
\end{equation}
Not very close to the transition ($\tau >\sqrt{{\rm Gi}}$) only the
excitations with energy $\omega >\Delta _{PG}$ are important. The pseudogap
does not play any role for such excitations. Thus, it is sufficient to
consider fluctuations in the linear approximation only (see [3], [4] and
[5]). It is important, however, that in the temperature region $\tau <\sqrt{%
{\rm Gi}}$ the excitations with energy $\omega <\Delta _{PG}$ become
essential. In [6] the fluctuation correction to conductivity was considered
in view of the pseudogap. The pseudogap was taken into account in the same
way as the gap below the transition temperature. This approximation gives a
correct estimate for the width of the temperature region where the
non-linear effects are important. However, the model considered in [6] with $%
\Delta $ being a constant can not reproduce the correct temperature
dependence for conductivity in the temperature 
region $\tau <\sqrt{{\rm Gi}}$%
.

To describe the nonlinear effects we will consider fluctuations of $\Delta$
in the statical approximation. This is eligible, since the lifetime of
fluctuations $(T \tau)^{-1}$ is large compared to the inverse pseudogap.
However, the spatial dispersion of the pseudogap changes the physical
picture significantly. To take into account the spatial variations, we have
to calculate the conductivity as a function of the order parameter $\Delta(%
{\bf r})$ being an arbitrary function of ${\bf r}$ and average out the
result over the Gaussian fluctuations with the correlator (\ref{1}). In the
present paper we accomplish this program up to a numerical coefficient in the
limiting case when the energy relaxation rate is large ($\tau_{\varepsilon}
\ll (T \tau)^{-1}$). In the other cases, we obtain a functional form of the
temperature dependence of conductivity without calculating coefficients.

To consider the spatial dependence of the order parameter we will
use the results obtained in [11]. In this work, it was shown that the
spatial variations of $\Delta$ act on one-particle excitations in the same
way as do the magnetic impurities. In this case, the total pairbreaking rate 
$\Gamma$ can be written as a sum of the pairbreaking rate due to the
magnetic impurities and the fluctuation term. Thus, the self-consistent
equation for $\Gamma$ takes the following form 
\begin{equation}  \label{3}
\Gamma = \int {\frac{d^2 k }{(2 \pi)^2}}\, \frac {\langle \Delta_k^*
\Delta_k \rangle} {\omega + {\frac{1 }{2}} D k^2 + \Gamma} + {\frac{1 }{%
\tau_s}}.
\end{equation}
It is important to mention, that Eq.(\ref{3}) is exact either if $\omega \gg
\Gamma$ or if $\tau_s$ is very small, so that the first term in Eq.(\ref{3})
is a small correction to the second one. In the other cases the
self-consistent Eq.(\ref{3}) can be considered as an estimate and gives the
result valid on the order of magnitude only.

In the region $\omega < \Gamma$, $\Gamma \gg T \tau$ we obtain from Eqs.(\ref
{3}), (\ref{1}): 
\begin{equation}  \label{4}
\Gamma = {\frac{8 T }{\pi}} \left( {\rm Gi} \, \ln{\frac{\Gamma }{T \tau}}
\right)^{1/2},
\end{equation}
which coincides with the value obtained in [7,12] up to the logarithmic
term. Below, we repeat the derivation done in [11] and show that the
pseudogap does not change result (\ref{4}) qualitatively.

Let us note, that the pair-breaking rate $\Gamma$ is of the order of the
pseudogap $\Delta_{PG}$. Thus, a wide maximum appears in the density of
states at $\omega \sim \Delta_{PG}$.

As it is known from [5], the Maki-Thomson correction to conductivity
saturates for $T \tau < \Gamma$ and takes the form: 
\begin{equation}  \label{5}
{\frac{\delta \sigma^{MT} }{\sigma_0}} = {\frac{8 T {\rm Gi} }{\pi \Gamma}}
\ln{\frac{\pi \Gamma }{4 T \tau}}.
\end{equation}
As it can be seen from Eqs.(\ref{4},\ref{5}) such a saturation takes place
when $\tau < \sqrt{{\rm Gi}}$. The similar results have been obtained in
[7,8,10]. However, numerical coefficients are different.

Note, that the numerical coefficient in Eq.(\ref{5}) depends on the way how
the summation of higher order diagrams is made. However, its exact value is
not very important since in the region $T \tau < \Gamma$ Maki-Thomson
contribution is less singular compared to Aslamazov-Larkin contribution and can
be neglected. Let us realize, that AL contribution does not saturate when 
$T$ tends to 
$T_c$ but becomes more and more singular.

To estimate AL contribution due to the appearance of the fluctuating Cooper
pairs one can use simple Drude formula: 
\begin{equation}
\delta \sigma ^{AL}={\frac{ne^{2}}{m}}\tau _{fl}.  \label{6}
\end{equation}
Where $n$, $m$ and $\tau _{fl}$ are concentration, mass and lifetime of the
fluctuating Cooper pairs. The ratio $n/m$ can be estimated from Eq.(\ref{1}%
), while the lifetime follows from the TDGL equations. 
\begin{equation}
\left( a{\frac{\partial }{\partial t}}+Dk^{2}+{\frac{8}{\pi }}T\tau \right)
\Delta _{k}(t)=\zeta (t),  \label{7}
\end{equation}
where $\zeta $ is the Langevin noise. In two-dimensional case we have 
\[
{\frac{n}{m}}\approx {\frac{T}{2\pi d\hbar ^{2}}}
\]
and 
\[
\tau _{fl}={\frac{\pi \hbar }{8(T-T_{c})a}}.
\]
Not very close to the transition ($T\tau >\Delta _{PG}$) or if the energy
relaxation rate is very large we can put $a=1$, since the quasiparticles are
in the thermal equilibrium. Thus, we have 
\begin{equation}
{\frac{\delta \sigma _{AL}}{\sigma }}={\frac{{\rm Gi}}{\tau }}.  \label{8}
\end{equation}

In the presence of the pseudogap there is no equilibrium and the coefficient 
$a$ becomes greater then one. Recall, that analogous changes in the
coefficient $a$ in TDGL equations for $|\Delta|$ appear below the transition
temperature (see e.g. [13,14,15,16,17]). The growth of the coefficient $a$
and, consequently, the growth of the fluctuations lifetime is due to that
the quasiparticles require more time to attain the thermal equilibrium 
(we denote the corresponding time  as $\tau_e$). 
A rough estimate gives 
$a \sim\Delta_{PG} \tau_e$. In the case of weak energy 
relaxation, $\tau_e$ has to
be determined from the diffusion equation in view of pseudogap (see
[18,19,20]). Note, that in this complicated case coefficient $a$ becomes a
non-local operator. Rough estimates give the following value for the thermal
equilibrium transition time $\tau_e \sim (D k_{{\rm min}}^2)^{-1} \sim (T
\tau)^{-1}$. Taking into account Eq.(\ref{2}) we obtain 
\begin{equation}  \label{9}
{\frac{\delta \sigma }{\sigma_0}} = {\frac{{\rm Gi}^{3/2} }{\tau^2}}.
\end{equation}
We see, that paraconductivity can exceed normal conductivity $\sigma_0$ in
the region ${\rm Gi}^{3/4} > \tau > {\rm Gi}$. 
Let us emphasize, that in this
region corrections to all the thermodynamic coefficients are small and the
linear theory is well applicable for them.

Let us now discuss the role of the energy relaxation processes, characterized by
a quasiparticle lifetime $\tau_\varepsilon$. Non-elastic electron-electron
scattering in dirty metals leads to the following electron-electron
collision times in 2D case 
\[
\tau_{\varepsilon}^{-1} \sim T d\, l\, p^2 \sim {\rm Gi} T. 
\]
Such a large collision time does not change nonlinear effects. However,
nonelastic electron scattering off phonons and other possible collective
excitations can decrease $\tau_{\varepsilon}$ significantly. These processes
together with additional pairbreaking processes (due to magnetic impurities
or magnetic field) lead to a decrease of the nonlinear effects. Energy
relaxation reduces the thermal equilibrium transition time $\tau_e$. If
these processes are very strong (for example if the temperature is
relatively large) transport equation for the distribution function becomes
local and in the limit $T\tau \sim D k^2 \ll \tau_{\varepsilon}^{-1}$ we can
write $\tau_e = \tau_{\varepsilon}$. Thus, in the temperature region under
consideration we have 
\begin{equation}  \label{10}
{\frac{\delta \sigma }{\sigma_0}} = {\frac{{\rm Gi}^{3/2} T
\tau_{\varepsilon} }{\tau}}.
\end{equation}
Elastic scattering off magnetic impurities and magnetic field also tend to
diminish the nonlinear fluctuation effects in conductivity but in a
different way. These scattering processes (as well as scatterings off the
static fluctuations of the order parameter) do not affect quasiparticle
motion and $\tau_{\varepsilon}$ thereby. However, if the pairbreaking rate
is large enough ($\Gamma > \Delta_{PG}$) these processes lead to the reduced
pseudogap $\Delta_{PG} \sim {\langle \left| \Delta \right|^2 \rangle / \Gamma%
}$ (recall, that without pairbreaking we would have $\Delta_{PG} \sim \sqrt{%
\langle \left| \Delta \right|^2 \rangle} \sim T {\rm Gi}^{1/2}$). Thus, the
fluctuation correction can be written as 
\begin{equation}  \label{11}
{\frac{\delta \sigma }{\sigma_0}} = {\frac{{\rm Gi}^{2} T }{\tau^2 \Gamma}}.
\end{equation}

In the presence of both a strong pairbreaking and a large energy relaxation
it is possible to derive exact expressions with logarithmic accuracy for the
coefficient $a$ in TDGL, which is local in this case, and for
paraconductivity. In this case, the main contribution to $a$ comes from the
fluctuations with $T \tau < D k^2 < \tau_{\varepsilon}^{-1}$. The first
inequality allows us to consider only the leading terms in the expansion $a$
over $\Delta$, while the second implies a local approximation in the
transport equation. The result is 
\begin{equation}  \label{12}
a = {\frac{\tau_{\varepsilon} \langle \Delta^2 \rangle }{2 \Gamma}},
\end{equation}
\begin{equation}  \label{13}
{\frac{\delta \sigma }{\sigma_0}} = {\frac{32 {\rm Gi}^2 T^2
\tau_{\varepsilon} }{\pi^2 \Gamma \tau}} \ln{\frac{\pi }{8 T
\tau_{\varepsilon} \tau}}.
\end{equation}
Note, that Eqs.(\ref{9}-\ref{13}) are valid only if the parameters $\Gamma$
and $\tau_{\varepsilon}$ are such that the contribution to conductivity $%
\delta \sigma$ is larger than the usual Aslamazov-Larkin contribution Eq.(\ref
{8}). If $\Gamma > T$, $T \tau_{\varepsilon} < \sqrt{{\rm Gi}}$ or if $T^2
\tau_{\varepsilon} / \Gamma < {\rm Gi}$, than nonlinear effects are
negligible and the usual result (\ref{8}) is valid for all $\tau> {\rm Gi}$.
Note, that MT contribution saturates at temperatures when $T \tau \sim {\rm max%
}\, \left[ \Gamma,\, {\frac{1 }{\tau_{\varepsilon}}},\, T \sqrt{{\rm Gi}}
\right]$.

\section*{2. Depairing factor induced by fluctuations}

Nonzero fluctuating order parameter $\Delta$ and Gor'kov green function $%
\beta$ [6] exist above the transition temperature. In the temperature region 
$\tau > {\rm Gi}$ the main contribution to the value of order parameter $%
\Delta$ arises from zero "frequency". The momentum space can be separated
into two parts: $\pi Dk^2/8T<\tau$ and $\pi Dk^2/8T>\tau$. The fluctuations
with $\pi Dk^2/8T>\tau$ can be considered as ``fast'' variables, created on
the background of slow fluctuations with $\pi Dk^2/8T<\tau$. The ``fast''
fluctuations induce intrinsic depairing factor $\Gamma$, even if external
depairing factor, connected with paramagnetic impurities is missing $%
(\tau_s\to\infty)$. Similar phenomenon was studied in paper [11]. Using the
method, developed in this paper, we obtain expressions for the statical
Green functions $\alpha, \beta$ and depairing factor $\Gamma$. We start from
Usadel equation for Green functions $\alpha$ and $\beta$ in the dirty limit
(see [6,21]): 
\begin{equation}  \label{14}
\Delta \alpha - \omega \beta +{\frac{D }{2}} \left( \alpha \nabla^2 \beta -
\beta \nabla^2 \alpha \right) = \alpha \beta \Gamma.
\end{equation}

According to paper [11] we present Green functions $\alpha, \beta$ in the
field of "fast" fluctuations of the order parameter $\Delta (k)$ in the form 
\begin{equation}  \label{15}
\alpha = \langle\alpha\rangle+\alpha_1, \quad \beta = \langle\beta\rangle
+\beta_1
\end{equation}

The deviations of the Green functions from their mean values can be found on
the perturbation theory. We have [11] 
\begin{equation}  \label{16}
\alpha_1(k)=- \frac{\Delta (k)\langle\alpha\rangle\langle\beta\rangle} {%
\omega\langle\alpha\rangle + \langle\Delta\rangle\langle\beta\rangle + Dk^2/2%
}
\end{equation}

The "mean" Green functions $\langle\alpha\rangle, \langle\beta\rangle$ are
solutions of the following system of equations: 
\begin{equation}  \label{17}
\langle\alpha\rangle^2+\langle\beta\rangle^2=1, \quad
\langle\alpha\rangle\langle\Delta\rangle - \omega\langle\beta\rangle =
\langle\alpha\rangle\langle\beta\rangle\Gamma
\end{equation}

The value of parameter $\Gamma$ is defined by Eq.(\ref{16}) and equals to 
\begin{equation}  \label{18}
\Gamma = \int\frac{d^2k}{(2\pi)^2}~ \frac{\langle\Delta_k^{*}\Delta_k\rangle%
} {\langle\alpha\rangle\omega +\langle\Delta\rangle \langle\beta\rangle +
Dk^2/2},
\end{equation}
where $\langle\Delta\rangle = \langle |\Delta |^2\rangle^{1/2}$. The
quantity $\langle\Delta\rangle$ in Eqs (\ref{16}), (\ref{17}) should be
understood as an integral over $k$ from the expression (\ref{18}) in the
range $\pi Dk^2/8T\leq\tau$ and takes the form 
\begin{equation}  \label{19}
\langle\Delta\rangle = \left [ \frac{T}{\nu d}\int\frac{d^2k}{(2\pi)^2} 
\frac{1}{\tau +\frac{\pi D}{8T}k^2}\right ]^{1/2}\approx T\left [\frac{64%
{\rm Gi}}{\pi^2}\right ]^{1/2}
\end{equation}
From Eqs(\ref{1}) and (\ref{18}) we obtain 
\begin{equation}  \label{20}
\Gamma = \frac{16T{\rm Gi}}{\pi\tau}~ \frac{1}{\frac{\pi}{4T\tau}%
(\omega\langle\alpha\rangle + \langle\Delta\rangle\langle\beta\rangle )-1}
\ln \left ( \frac{\pi(\omega\langle\alpha\rangle +
\langle\Delta\rangle\langle\beta\rangle )} {4T\tau}\right )
\end{equation}

As  can be seen from Eq.(\ref{20}), $\Gamma(\omega)$ is a function of
energy $\omega$. In the range $\tau\leq\sqrt{{\rm Gi}}$ essential values of $%
\omega$ are of the order of $\Gamma$. Thus, $\Gamma$ itself is of the order
of $\langle\Delta\rangle$ (see (19)). This order of $\Gamma$ is connected
with fluctuations of the modulus of the order parameter. This value is much
larger, then the one due to the phase fluctuations of the order parameter
(see [6]).

\section*{3. Equations for the time dependent order parameter}

Statical Ginzburg-Landau equations are valid in wide temperature region 
\begin{equation}  \label{21}
{\rm Gi}\ll |1-T/T_c|\ll 1
\end{equation}

TDGL equations are valid if the energy relaxation time $\tau_{\varepsilon}$
or the pair breaking time $\tau_s=\Gamma^{-1}$ is short enough [13-16]. For
large $\tau_{\varepsilon}$'s the dynamics of normal excitations becomes
essential. As a result, the dynamical term in the equation for the order
parameter becomes more complicated. Below, we derive the corresponding
equation.

Order parameters $\Delta_{1,2}(t)$ can be written as 
\begin{equation}  \label{22}
\Delta_{1,2}(t)=\frac{\pi\lambda_{eff}}{2} {\it F}^K_{1,2}(t,t)
\end{equation}
where %{\it F}^K_{1,2}$ is one of Green functions. The general 
Green function $\hat G$ can be presented in the form [18] 
\begin{equation}  \label{23}
\hat G=\left ( 
\begin{array}{ll}
G^R; & G^K \\ 
0; & G^A
\end{array}
\right )
\end{equation}
where $G^{R,A,K}$ are retarded, advanced and Keldish Green functions. Each
of them is matrix of Gor'kov-Nambu 
\begin{equation}  \label{24}
G^{R,A,K}=\left ( 
\begin{array}{ll}
g_1; & {\it F}_1 \\ 
-{\it F}_2; & g_2
\end{array}
\right )^{R,A,K}, \quad\quad \tilde\Delta =\left ( 
\begin{array}{cc}
0; & \Delta_1 \\ 
-\Delta_2; & 0
\end{array}
\right )
\end{equation}
where $\Delta_2(\omega)=\Delta^{*}_1(-\omega)$.

In the dirty limit we have the following system of equations for $G^{R,A}$
(see [19]) 
\begin{equation}  \label{25}
-D\partial_{\mp}\left ( g^{R,A}\partial_{\mp}{\it F}^{R,A}_{1,2}- {\it F}%
^{R,A}_{1,2}\frac{\partial g^{R,A}}{\partial r} \right )
+2i\Delta_{1,2}g^{R,A}-
\end{equation}
\[
-2i\varepsilon{\it F}^{R,A}_{1,2} +\frac{2}{\tau_s}g^{R,A}{\it F}%
^{R,A}_{1,2}=-I^{Ph(R,A)}_{1,2} 
\]
where the electron-phonon collision integral $I^{Ph(R,A)}_{1,2}$ in the
vicinity of the transition temperature $T_c$ for small values of energy $%
|\varepsilon |\ll T$ is equal to 
\begin{equation}  \label{26}
I^{Ph(R,A)}_{1,2}=\pm\frac{1}{\tau_{\varepsilon}} {\it F}^{R,A}_{1,2}
\end{equation}

The Keldysh Green function $G^K$ can be presented in the form [20] 
\begin{equation}  \label{27}
G=\int dt_1(G^R\hat f-\hat fG^A)
\end{equation}
where distribution function $\hat f$ is given as [20] 
\begin{equation}  \label{28}
\hat f=f+\tau_zf_1
\end{equation}

Equations for the distribution functions $f_{1,2}$ has been derived in paper
[20] and have the form 
\begin{equation}  \label{29}
-D\frac{\partial}{\partial r}\Biggl \{ \frac{\partial f}{\partial r}%
(1-G^RG^A) \Biggr \}- D\frac{\partial}{\partial r}(f_1j_{\varepsilon})+ 2%
\frac{\partial f}{\partial t} Sp\alpha +
\end{equation}
\[
+\frac{\partial f}{\partial\varepsilon} \Biggl \{ eD\frac{\partial A}{%
\partial t}j_{\varepsilon}- 2Sp\frac{\partial\hat\Delta}{\partial t}\delta %
\Biggr \}+ 4I^{Ph}_1(f)=0, 
\]
\[
-D\frac{\partial}{\partial r} Sp\Biggl \{
\frac{\partial f_1}{\partial r}(1-\tau_zG^R\tau_zG^A) \Biggr \}- D\frac{%
\partial f}{\partial r}j_{\varepsilon}+ 2\frac{\partial}{\partial t}%
(f_1Sp\alpha)- 
\]
\[
-4if_1Sp(\gamma\hat\Delta)+2\frac{\partial f}{\partial\varepsilon}Sp \Biggl
\{ e\frac{\partial\varphi}{\partial t}\alpha - \frac{\partial\hat\Delta}{%
\partial t}\tau_z\gamma+ \frac{i}{2}\frac{\partial^2\hat\Delta}{\partial t^2}
\frac{\partial\delta}{\partial\varepsilon}\Biggr \} + 
\]
\[
+4I^{Ph}_2(f_1)=0. 
\]
where 
\begin{equation}  \label{30}
j_{\varepsilon}=Sp\tau_z(G^R\partial G^R-G^A\partial G^A), \quad\quad
\partial =\frac{\partial}{\partial r}-ieA\tau_z,
\end{equation}
\[
2\alpha = G^R\tau_z-\tau_zG^A, \quad\quad 2\delta = G^R-G^A, \quad\quad
2\gamma = G^R+G^A 
\]

In the important limiting case $\varepsilon\sim\Gamma\gg\Delta$ the Eqs.(\ref
{25},\ref{29}) can be simplified and we obtain 
\begin{equation}  \label{31}
{\it F}^{R,A}_1=-i\Biggl ( \Gamma\mp i\varepsilon - \frac{D}{2}\frac{%
\partial^2}{\partial r^2} \Biggr )^{-1}\Delta;
\end{equation}
\[
{\it F}^{R,A}_2=-i\Biggl ( \Gamma\mp i\varepsilon - \frac{D}{2}\frac{%
\partial^2}{\partial r^2} \Biggr )^{-1}\Delta^{*}; 
\]
\[
-D\frac{\partial^2f}{\partial r^2}- \frac{D}{4}\frac{\partial}{\partial r}%
(j_{\varepsilon}f_1)+ \frac{\partial f}{\partial t}+\frac{1}{4} \frac{%
\partial f}{\partial\varepsilon} \Biggl \{
\frac{\partial\Delta_1}{\partial t}({\it F}^R_2-{\it F}^A_2)+ 
\]
\[
+\frac{\partial\Delta_2}{\partial t}({\it F}^R_1-{\it F}^A_1) \Biggr \}
+I^{Ph}_1(f)=0 
\]
\[
-D\frac{\partial^2f_1}{\partial r^2}- \frac{D}{4}j_{\varepsilon}\frac{%
\partial f_1}{\partial r}+ \frac{\partial f_1}{\partial t}+ \frac{i}{2}f_1
\left (\Delta ({\it F}^R_2+{\it F}^A_2)+ \Delta^{*}({\it F}^R_1+{\it F}%
^A_1)\right )+ 
\]
\[
\frac{\partial f}{\partial\varepsilon} \Biggl \{
e \frac{\partial\varphi}{\partial t} +\frac{1}{4}\left ( - \frac{%
\partial\Delta_1}{\partial t}({\it F}^R_2+{\it F}^A_2)+ \frac{%
\partial\Delta_2}{\partial t}({\it F}^R_1+{\it F}^A_1) \right )\Biggr \} 
+I^{Ph}_2(f_1)=0, 
\]
where 
\begin{equation}  \label{19}
j_{\varepsilon}=- {\it F}^R_1\frac{\partial{\it F}^R_2}{\partial r}+ {\it F}%
^R_2\frac{\partial{\it F}^R_1}{\partial r}+ {\it F}^A_1\frac{\partial{\it F}%
^A_2}{\partial r}- {\it F}^A_2\frac{\partial{\it F}^A_1}{\partial r}
\end{equation}

The general expression for collision integrals $I^{Ph}_{1,2}$ is given in
papers [20,22]. For small values of energy $|\varepsilon |\ll T$ these
integrals can be taken in the simple form 
\begin{equation}  \label{33}
I^{Ph}_1(f)=\frac{1}{\tau_{\varepsilon}} \left ( -\mbox {\rm th} \left (%
\frac{\varepsilon}{2T} \right )+f \right )
\end{equation}
\[
I^{Ph}_2(f_1)=\frac{1}{\tau_{\varepsilon}} f_1 
\]
\[
\tau^{-1}_{\varepsilon}=7 \zeta(3) \pi\nu g^2T^3/2(s p)^2 
\]
where $s$ is the velocity of sound in metal and $g$ is the electron-phonon
coupling constant.

In the limiting case of strong energy relaxation, when $\tau_{\varepsilon}%
\Delta\ll 1$ the distribution function $\hat f$ can be taken as the
equilibrium one 
\begin{equation}  \label{34}
f=\mbox {\rm th}(\varepsilon /2T), \quad\quad f_1=0
\end{equation}

In this case one can obtain from Eqs.(\ref{22}), (\ref{31}) and (\ref{34})
time-dependent Ginzburg-Landau equation in the usual form 
\begin{equation}  \label{35}
\left ( 1-T/T_c - \frac{7\zeta (3)} {8\pi^2T^2}|\Delta |^2 \right )\Delta + 
\frac{\pi D}{8T}\partial^2_{-}\Delta - \frac{\pi}{8T} \left ( \frac{\partial%
}{\partial t}+2ie\varphi \right ) \Delta =0
\end{equation}

If condition $\tau_{\varepsilon}\Delta \ll 1$ is not satisfied, then the
deviation of the distribution function $\hat f$ from it's equilibrium value
can lead to the change of last term in Eq.(\ref{35}).

In the range $\Gamma \gg \Delta$ the crossing term in Eq.(\ref{31}) is small
on the parameter $(\Delta/\Gamma )^2$. Thus, in the leading approximation
system (31) is diagonal.

With the aid of Eqs.(\ref{27}), (\ref{28}) and (\ref{31}) we can rewrite Eq.(%
\ref{22}) in the form 
\begin{equation}  \label{36}
\left [ \tau + \frac{\pi}{8T} \left ( -i\omega_1 - D\frac{\partial^2}{%
\partial r^2} \right ) \right ] \Delta_1 - \frac{\pi}{2}\int\limits^{%
\infty}_{-\infty} \frac{d\varepsilon}{2\pi} \left [ \delta f ({\it F}^R_1-%
{\it F}^A_1)-f_1({\it F}^R_1+{\it F}^A_1) \right ] =0
\end{equation}
\[
\left [ \tau + \frac{\pi}{8T} \left ( -i\omega_1 - D\frac{\partial^2}{%
\partial r^2} \right ) \right ] \Delta_2 - \frac{\pi}{2}\int\limits^{%
\infty}_{-\infty} \frac{d\varepsilon}{2\pi} \left [ \delta f ({\it F}^R_2-%
{\it F}^A_2)-f_1({\it F}^R_2+{\it F}^A_2) \right ] =0 
\]
In Eqs.(\ref{36}) we put 
\begin{equation}  \label{37}
f=\mbox {\rm th} (\varepsilon /2T)+\delta f
\end{equation}

In (\ref{36}) the contributions due to the second terms are of the order $%
(\Delta /\Gamma)^2$. This result is due to the cancellation of terms, going
from quantities $\delta f$ and $f_1$. But in the next orders of the perturbation
theory the quantity $f_1$ becomes small and the main contribution arises
from distribution function $\delta f$ beyond the perturbation theory.

\section*{4. The conductivity of fluctuating pairs \newline
(Aslamazov-Larkin contribution)}

The conductivity of fluctuating pairs is given by diagrams on Fig. 1a. 
We
suppose below, that order parameters $\Delta _{1,2}$ can be written as a sum
of two terms. One of them is connected with the statical thermodynamic
fluctuations $\Delta ,\Delta ^{*}$. In the range $\tau >{\rm Gi}$ these
fluctuations are Gaussian with the correlator given by Eq.(\ref{1}). The
wavy line on Fig. 1a gives the dynamical fluctuations $\tilde{\Delta}_{1,2}$
of the order parameter. The correlators of these fluctuations $\hat{K}_{ij}$
should be found on the background of thermodynamic fluctuations 
\begin{equation}
\hat{K}_{ij}(\omega _{1})=\nu \langle \Delta _{i}^{*}\Delta _{j}\rangle
_{\omega _{1}}  \label{38}
\end{equation}

The contribution to conductivity can be expressed through correlators $\hat K$
in the same way as in the case of weak fluctuations [3].

First, we have to find conductivity as a function of Matzubara frequency $%
\omega _{0}$ and, then, perform analytical continuation on $\omega _{0}$.
The correction to current was found in paper [6] with the aid of the
equations for Green function in the dirty limit in high frequency fields 
\begin{equation}
j_{\omega _{0}}^{\alpha }=\frac{1}{2d}\int d^{2}r_{1}T\sum\limits_{\omega
_{1}}Sp\hat{L}_{r}^{\alpha }\hat{K}(\omega _{1}+\omega _{0},r,r_{1})\hat{L}%
_{r_{1}}^{\beta }\hat{K}(\omega _{1},r_{1},r)A_{\omega _{0}}^{\beta }
\label{39}
\end{equation}
where $A_{\omega _{0}}$is the vector potential of the external field, and
matrix $\hat{L}$ is equal to 
\begin{equation}
\hat{L}_{12}^{\alpha }=L_{21}^{\alpha }=0,\quad \hat{L}_{11}^{\alpha }(r)=-%
\frac{\pi eD}{2T}\frac{\partial }{\partial r^{\alpha }},\quad \hat{L}%
_{22}^{\alpha }=-\hat{L}_{11}^{\alpha }  \label{40}
\end{equation}

\begin{figure}
\epsfxsize=5in
\centerline{\epsfbox{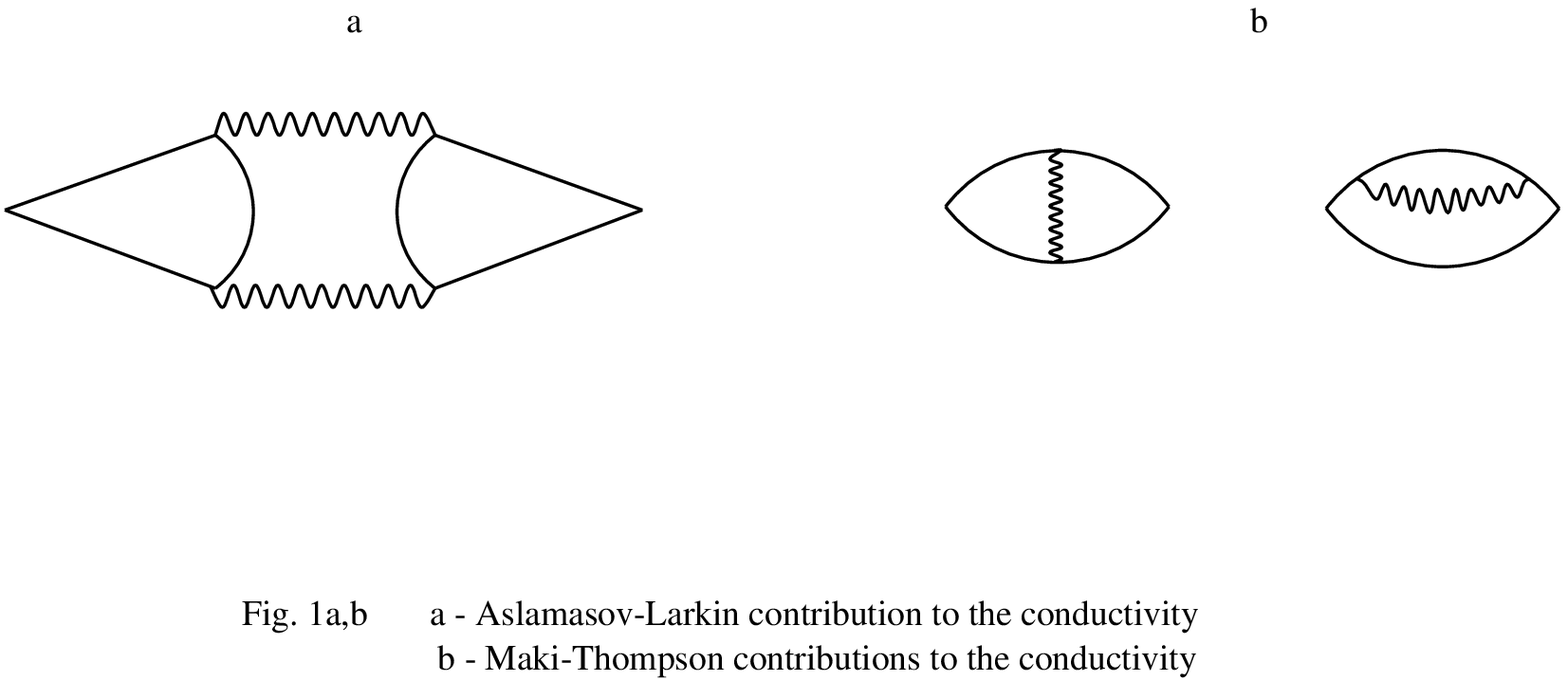}}
\end{figure}

After the analytical continuation over $\omega _{0}$ in Eq.(\ref{39}) we
obtain 
\[
j_{\omega }^{\alpha }=-\frac{1}{2d}\int d^{2}r_{1}\frac{iT}{2\pi }%
\int\limits_{-i\infty }^{i\infty }d\omega _{1}\Biggl [ \frac{1}{\omega
_{1}-i\omega -\delta }-\frac{1}{\omega _{1}+\delta }\Biggr ] \cdot 
\]
\begin{equation}
Sp\left( \hat{L}_{r}^{\alpha }\hat{K}(\omega _{1}-i\omega +\delta ,r,r_{1})%
\hat{L}_{r_{1}}^{\beta }\hat{K}(\omega _{1}-\delta ,r_{1},r)\right)
A_{\omega }^{\beta }  \label{41}
\end{equation}

It was found in paper [6], that fluctuations are weak in the range $\tau >(%
{\rm Gi})^{1/2}$. And in this region we have 
\begin{equation}  \label{42}
K_{11}(\omega_1+\delta)=K_{22}(\omega_1+\delta)= 
\frac{1}{\tau +\frac{\pi}{8T%
}(\omega_1+Dk^2)}
\end{equation}
From Eqs.(\ref{40}--\ref{42}) we obtain the well known result for the
paraconductivity [3] 
\begin{equation}  \label{43}
\sigma^{(a)}/\sigma_0=\frac{{\rm Gi}}{\tau}
\end{equation}

To obtain the conductivity in the temperature region $\tau < ({\rm Gi})^{1/2}
$ we should find correlation functions $\hat K$ in the field of
thermodynamic fluctuations $\Delta$. Than we have to average out over $\Delta
$ the expression for conductivity. The correlation functions $\hat K$ can be
found from Eq.(\ref{36}). 
\begin{equation}  \label{44}
\hat K^{-1}= \Bigg ( 
\begin{array}{ll}
\tau +\frac{\pi}{8T}\left ( \omega_1-D\frac{\partial^2}{\partial r^2} \right
) -C_{11}; & -C_{12} \\ 
-C_{21}; & \tau + \frac{\pi}{8T}\left ( \omega_1-D\frac{\partial^2}{\partial
r^2} \right )-C_{22}
\end{array}
\Biggr )
\end{equation}
where operators $C_{ij}$ are given by 
\begin{equation}  \label{45}
C_{11}=\frac{\pi}{2}\int\limits^{\infty}_{-\infty} \frac{d\varepsilon}{2\pi}%
\Biggl [ \left ( {\it F}^R_1-{\it F}^A_1 \right ) \delta
f^{(1)}-f^{(1)}_1\left ( {\it F}^R_1+{\it F}^A_1 \right ) \Biggr ],
\end{equation}
\[
C_{12}=\frac{\pi}{2}\int\limits^{\infty}_{-\infty} \frac{d\varepsilon}{2\pi}%
\Biggl [ \left ( {\it F}^R_1-{\it F}^A_1 \right ) \delta
f^{(2)}-f^{(2)}_1\left ( {\it F}^R_1+{\it F}^A_1 \right ) \Biggr ], 
\]
\[
C_{21}=\frac{\pi}{2}\int\limits^{\infty}_{-\infty} \frac{d\varepsilon}{2\pi}%
\Biggl [ \left ( {\it F}^R_2-{\it F}^A_2 \right ) \delta f^{(1)}+\left ( 
{\it F}^R_2+{\it F}^A_2 \right ) \delta f^{(1)}_1 \Biggr ], 
\]
\[
C_{22}=\frac{\pi}{2}\int\limits^{\infty}_{-\infty} \frac{d\varepsilon}{2\pi}%
\Biggl [ \left ( {\it F}^R_2-{\it F}^A_2 \right ) \delta f^{(2)}+\left ( 
{\it F}^R_2+{\it F}^A_2 \right ) \delta f^{(2)}_1 \Biggr ]
\]

In equations (45) the operators $\delta f^{(1,2)},f_{1}^{(1,2)}$ are defined
as 
\begin{equation}
\delta f=\delta f^{(1)}\tilde{\Delta}_{1}+\delta f^{(2)}\tilde{\Delta}_{2},
\label{46}
\end{equation}
\[
f_{1}=f_{1}^{(1)}\tilde{\Delta}_{1}+f_{1}^{(2)}\tilde{\Delta}_{2}
\]
and $\delta f,f_{1}$ are solutions of the system (31) in field of $\tilde{%
\Delta}_{1},\tilde{\Delta}_{2}$. For an arbitrary function $\Delta (r)$
system (31) can not be solved analytically.. Nevertheless, in the range $%
\tau <({\rm Gi})^{1/2}$ the expression for correlation functions $\hat{K}$
can be found with the logarithmic accuracy if the value of the external
depairing factor $\Gamma $ is larger then $\Delta $. In this case simple
expression for Green functions ${\it F}_{1,2}^{R,A}$ 
can be used 
\begin{equation}
{\it F}_{1}^{R,A}=\frac{-i\Delta }{\Gamma \mp i\varepsilon },\quad \quad 
{\it F}_{2}^{R,A}=\frac{-i\Delta ^{*}}{\Gamma \mp i\varepsilon }  \label{47}
\end{equation}

If $Dk^2\gg |\Delta |^2/\Gamma$, the contribution of $\delta f^{(1,2)}$ is
cancelled out in quantities $C_{11}$ and $C_{22}$. Note, that if $Dk^2 \ll
|\Delta |^2/\Gamma$, than $f^{(1,2)}_1\ll \delta f^{(1,2)}$. Thus, this
region gives the dominant contribution to $C_{ij}$. Eqs.(\ref{44},\ref{45})
in this case can be reduced to the following form 
\begin{equation}  \label{48}
\Biggl [ \tau +\frac{\pi}{8T}\Bigl ( \omega_1-D\frac{\partial^2} {\partial
r^2} \Bigr ) \Biggr ] K_{11}+ \frac{\pi\omega_1\Delta}{16T\Gamma} \Bigl (
\omega_1+\tau^{-1}_{\varepsilon}- D\frac{\partial^2}{\partial r^2} \Bigr
)^{-1}\Bigl ( \Delta^{*}K_{11} \Bigr )+
\end{equation}
\[
+\frac{\pi\omega_1\Delta}{16T\Gamma} \Bigl (
\omega_1+\tau^{-1}_{\varepsilon}- D\frac{\partial^2}{\partial r^2} \Bigr
)^{-1}\Bigl ( \Delta K_{21} \Bigr )=\delta (r-r_1), 
\]
\[
\Biggl [ \tau +\frac{\pi}{8T}\Bigl ( \omega_1-D\frac{\partial^2} {\partial
r^2} \Bigr ) \Biggr ] K_{12}+ \frac{\pi\omega_1\Delta}{16T\Gamma} \Bigl (
\omega_1+\tau^{-1}_{\varepsilon}- D\frac{\partial^2}{\partial r^2} \Bigr
)^{-1}\Bigl ( \Delta^{*}K_{12} \Bigr )+ 
\]
\[
+\frac{\pi\omega_1\Delta}{16T\Gamma} \Bigl (
\omega_1+\tau^{-1}_{\varepsilon}- D\frac{\partial^2}{\partial r^2} \Bigr
)^{-1}\Bigl ( \Delta K_{22} \Bigr )=0, 
\]
\[
\Biggl [ \tau +\frac{\pi}{8T}\Bigl ( \omega_1-D\frac{\partial^2} {\partial
r^2} \Bigr ) \Biggr ] K_{21}+ \frac{\pi\omega_1\Delta^{*}}{16T\Gamma} \Bigl
( \omega_1+\tau^{-1}_{\varepsilon}- D\frac{\partial^2}{\partial r^2} \Bigr
)^{-1}\Bigl ( \Delta K_{21} \Bigr )+ 
\]
\[
+\frac{\pi\omega_1\Delta^{*}}{16T\Gamma} \Bigl (
\omega_1+\tau^{-1}_{\varepsilon}- D\frac{\partial^2}{\partial r^2} \Bigr
)^{-1}\Bigl ( \Delta^{*} K_{11} \Bigr )=0, 
\]
\[
\Biggl [ \tau +\frac{\pi}{8T}\Bigl ( \omega_1-D\frac{\partial^2} {\partial
r^2} \Bigr ) \Biggr ] K_{22}+ \frac{\pi\omega_1\Delta^{*}}{16T\Gamma} \Bigl
( \omega_1+\tau^{-1}_{\varepsilon}- D\frac{\partial^2}{\partial r^2} \Bigr
)^{-1}\Bigl ( \Delta K_{22} \Bigr )+ 
\]
\[
+\frac{\pi\omega_1\Delta^{*}}{16T\Gamma} \Bigl (
\omega_1+\tau^{-1}_{\varepsilon}- D\frac{\partial^2}{\partial r^2} \Bigr
)^{-1}\Bigl ( \Delta^{*} K_{12} \Bigr )=\delta (r-r_1), 
\]

This system can be solved with logarithmic accuracy in the case of strong
energy relaxation $\tau^{-1}_{\varepsilon}> T\tau$. In this region we obtain
from Eq.(\ref{1}) 
\begin{equation}  \label{49}
\langle\Delta^{*}\Bigl (\omega_1+\tau^{-1}_{\varepsilon}- D\frac{\partial^2}{%
\partial r^2}\Bigr )^{-1} \Delta\rangle = \frac{64{\rm Gi}}{\pi^2}%
T^2\tau_{\varepsilon}\ln \Biggl ( \frac{\pi}{8T\tau\tau_{\varepsilon}} %
\Biggr )
\end{equation}

From Eqs. (\ref{1}), (\ref{48}) and (\ref{49}) we obtain the following
expression for the correlators $\hat K$: 
\begin{equation}  \label{50}
\Biggl \{ \tau +\frac{\pi D}{8T}k^2+ \frac{4{\rm Gi}T\tau_{\varepsilon}%
\omega_1}{\pi\Gamma} \ln \Biggl ( \frac{\pi}{8T\tau\tau_{\varepsilon}} %
\Biggr ) -
\end{equation}
\[
\frac{2}{\tau} \Biggl ( \frac{4{\rm Gi}T\tau_{\varepsilon}\omega_1} {%
\pi\Gamma} \Biggr )^2 I \Biggr \} K_{11}=1, \quad\quad\quad K_{22}=K_{11} 
\]
where 
\begin{equation}  \label{51}
I=\int\limits^{\infty}_0 \frac{dxdy}{(x+1)(y+1)\sqrt{(x-y)^2+2(x+y)a+a^2}},
\end{equation}
\[
a=1+\frac{4\omega_1T\tau_{\varepsilon}{\rm Gi}}{\pi\Gamma\tau} \ln \Biggl ( 
\frac{\pi}{8T\tau\tau_{\varepsilon}} \Biggr )
\]

The nondiagonal elements in $\hat K$ give logarithmically small contribution
to conductivity. As a result we obtain 
\begin{equation}  \label{52}
\frac{\sigma^a}{\sigma_0}= \frac{32{\rm Gi}^2T^2\tau_{\varepsilon}}{%
\pi^2\Gamma\tau} \ln\Biggl ( \frac{\pi}{8T\tau\tau_{\varepsilon}} \Biggr )
\end{equation}

The situation becomes more complicated if energy relaxation time $%
\tau_{\varepsilon}$ is large. From (\ref{48}) we obtain the following
equation for correlator $K_{11}$ 
\begin{equation}  \label{53}
\Biggl [ \tau +\frac{\pi}{8T}\Bigl ( \omega_1-D\frac{\partial^2} {\partial
r^2} \Bigr ) \Biggr ] K_{11}+ \frac{\pi\omega_1\Delta}{16T\Gamma} \Bigl (
\omega_1+\tau^{-1}_{\varepsilon}- D\frac{\partial^2}{\partial r^2} \Bigr
)^{-1}\Bigl ( \Delta^{*}K_{11} \Bigr )-
\end{equation}
\[
-\Biggl ( \frac{\pi\omega_1}{16T\Gamma} \Biggr )^2 \Delta \Bigl (
\omega_1+\tau^{-1}_{\varepsilon}- D\frac{\partial^2}{\partial r^2} \Bigr
)^{-1} \Delta \Biggl [ \tau + \frac{\pi}{8T} \Bigl ( \omega_1-D\frac{%
\partial^2}{\partial r^2} \Bigr ) + 
\]
\[
+\frac{\pi\omega_1}{16T\Gamma}\Delta^{*} \Bigl (
\omega_1+\tau^{-1}_{\varepsilon}- D\frac{\partial^2}{\partial r^2} \Bigr
)^{-1}\Delta \Biggr ]^{-1} \Delta^{*} \Bigl ( \omega_1
+\tau^{-1}_{\varepsilon}- D\frac{\partial^2}{\partial r^2} \Bigr )^{-1}
\Bigl ( \Delta^{*}K_{11} \Bigr ) = 
\]
\[
=\delta (r-r_1) 
\]

First, we find the mean value for the following quantity 
\begin{equation}  \label{54}
\langle\Delta^{*}\Bigl (\omega_1- D\frac{\partial^2}{\partial r^2}\Bigr
)^{-1} \Bigl ( \Delta\exp (ikr) \Bigr ) \rangle =
\end{equation}
\[
\frac{64T^2{\rm Gi}}{\pi^2} \frac{1}{Dk^2+\frac{8T\tau}{\pi}} \ln \Biggl (
\frac{\pi (Dk^2+\frac{8T\tau}{\pi})^2}{8T\tau\omega_1} \Biggr )
\]

From here we see that the coefficient near $\omega _{1}$ in equation for $%
K_{11}$ is logarithmically large. Contrary to the previous case $(\tau
_{\varepsilon }^{-1}\gg T\tau )$ the last term in the right-hand side of Eq.(%
\ref{53}) is essential and leads together with nondiagonal elements in $\hat{%
K}$ to the cancellation of large terms in conductivity. To check this, we
have to find the mean value of the product of four $\Delta $ in last term in
Eq.(\ref{53}). We have 
\begin{equation}
I_{1}=\Biggl ( \frac{\pi }{16T\Gamma }\Biggr )^{2}\langle \Delta \left(
\omega _{1}-D\frac{\partial ^{2}}{\partial r^{2}}\right) ^{-1}\Delta \Biggl
[ \tau +\frac{\pi }{8T}\left( \omega _{1}-D\frac{\partial ^{2}}{\partial
r^{2}}\right) +  \label{55}
\end{equation}
\[
\frac{\pi \omega _{1}}{16T\Gamma }\Delta ^{*}\left( \omega _{1}-D\frac{%
\partial ^{2}}{\partial r^{2}}\right) ^{-1}\Delta \Biggr ]^{-1}\Delta
^{*}\left( \omega _{1}-D\frac{\partial ^{2}}{\partial r^{2}}\right)
^{-1}\Delta ^{*}\exp (ikr)\rangle =
\]
\[
=\exp (ikr)\Biggl ( \frac{\pi }{16T\Gamma \nu d}\Biggr )^{2}\int \frac{%
d^{2}k_{1}}{(2\pi )^{2}}\int \frac{d^{2}k_{2}}{(2\pi )^{2}}~\left[ \left(
\tau +\frac{\pi D}{8T}k_{1}^{2}\right) \left( \tau +\frac{\pi D}{8T}%
k_{2}^{2}\right) \right] ^{-1}\cdot 
\]
\[
\cdot \left[ (\omega _{1}+D(k-k_{2})^{2})(\omega _{1}+D(k-k_{1})^{2})\left(
\tau +\frac{\pi }{8T}Dk_{3}^{2}+\omega _{1}\alpha _{k_{3}}\right) \right]
^{-1}
\]
where 
\begin{equation}
k_{3}=k-k_{1}-k_{2},\quad \quad \alpha _{k}=\frac{4T{\rm Gi}}{\pi \Gamma }%
\frac{1}{Dk^{2}+\frac{8T\tau }{\pi }}\ln \Biggl ( \frac{\pi \left( Dk^{2}+%
\frac{8T\tau }{\pi }\right) ^{2}}{8T\tau \omega _{1}}\Biggr )  \label{56}
\end{equation}

The $\ln^2$ term can be easy separated from expression (\ref{55}). As result
we obtain 
\begin{equation}  \label{57}
I_1=\frac{1}{\tau +\frac{\pi D}{8T}k^2+\omega_1\alpha_k} \Biggl \{\alpha^2_k-%
\frac{4\pi\alpha_k{\rm Gi}}{\Gamma} \cdot
\end{equation}
\[
\cdot \int \frac{d^2k_1}{(2\pi)^2}~ \frac{\frac{\pi D}{8T}%
(k^2_1-k^2)+\omega_1(\alpha_{k_1}-\alpha_k)} {\left (\tau +\frac{\pi D}{8T}%
k^2_1\right )(k_1-k)^2 \left (\tau +\frac{\pi D}{8T}(k_1-k)^2+
\omega_1\alpha_{k_1-k}\right ) } \Biggr \}
\]

In Eq.(\ref{55}) we omitted the ''diagonal'' term with denominator of type $%
[\omega _{1}+D(k+k_{1})^{2}]^{2}$. This term leads to a small correction to
coefficient near $\omega _{1}$ in (\ref{53}).

With the same accuracy we present here the expression for nondiagonal
elements $K_{12},K_{21}$ as 
\begin{equation}  \label{58}
K_{21}=-\frac{\pi\omega_1}{16T\Gamma} \int \frac{d^2k_1d^2k_2}{(2\pi)^4}
\cdot
\end{equation}
\[
\cdot \frac{\Delta^{*}_{k_1}\Delta^{*}_{k_2}K_{11}(k)} {(\omega_1+D(k-k_1)^2)%
\left [ \tau + \frac{\pi D}{8T}(k-k_1-k_2)^2+ \omega_1\alpha_{k-k_1-k_2}
\right ]}, 
\]
\[
K_{12}=-\frac{\pi\omega_1}{16T\Gamma} \int \frac{d^2k_3d^2k_4}{(2\pi)^4}
\cdot 
\]
\[
\cdot \frac{\Delta_{k_3}\Delta_{k_4}K_{22}(k)} {(\omega_1+D(k+k_3)^2)\left [
\tau + \frac{\pi D}{8T}(k+k_3+k_4)^2+ \omega_1\alpha_{k+k_3+k_4} \right ]} 
\]

With the aid of Eqs.(\ref{57},\ref{58}) we obtain that correction to
conductivity in the form 
\begin{equation}  \label{59}
\frac{\sigma^a}{\sigma_0}\approx \frac{4T{\rm Gi}^2}{\pi\Gamma\tau^2}
\end{equation}

This expression is valid up to a numerical factor of the order of unity.

If the external depairing factor $\Gamma$ is zero (superconductor without
paramagnetic impurities), then the quantity $\Gamma$ in Eqs.(\ref{51},\ref
{59}) should be replaced by its intrinsic value 
\begin{equation}  \label{60}
\Gamma\approx T{\rm Gi}^{1/2}
\end{equation}
(see Eq.(\ref{18})).

As a result we obtain in the temperature region ${\rm Gi}<\tau <({\rm Gi}%
)^{1/2}$ 
\begin{equation}  \label{61}
\frac{\sigma^a}{\sigma_0}\approx\frac{4{\rm Gi}^{3/2}}{\pi\tau^2}
\end{equation}

Eq.(\ref{61}) means, that A.L. contribution to conductivity is strongly
enhanced in the temperature region ${\rm Gi}<\tau <({\rm Gi})^{1/2}$.

\section*{5. Maki-Thompson contribution to conductivity in nonlinear on
fluctuations region}

The general expression for Maki-Thompson contribution to conductivity $(\sigma
^{b})$ was given in paper [6]. Equation (28) in [6] can be considered as
interpolation expression for Maki-Thompson contribution to the conductivity,
that is valid in the entire temperature region $\tau >{\rm Gi}$. The
depairing factor $\Gamma $ in Eq.(28) in paper [6] should be changed to be a
sum of two terms: the first one is the external depairing factor $\tau
_{s}^{-1}$, connected with the spin flip scattering on magnetic impurities,
and the second one is the intrinsic depairing factor, given by Eq.(\ref{20}%
). As a result we obtain 
\begin{equation}
\frac{\sigma ^{b}}{\sigma _{0}}=\frac{\pi }{8d\nu }\int \frac{d^{2}k}{(2\pi
)^{2}}~\frac{1}{\Gamma +Dk^{2}/2}~\frac{1}{\tau +\frac{\pi D}{8T}k^{2}}=
\label{62}
\end{equation}
\[
=\frac{2{\rm Gi}}{\tau }~\frac{1}{\frac{\pi \Gamma }{4T\tau }-1}\ln \Biggl ( 
\frac{\pi \Gamma }{4T\tau }\Biggr )
\]

In the range ${\rm Gi} < \tau < ({\rm Gi})^{1/2}$ Maki-Thompson contribution
reaches its saturation value and effectively becomes temperature independent 
\begin{equation}  \label{63}
\frac{\sigma^b}{\sigma_0}=({\rm Gi})^{1/2}\ln \Biggl ( \frac{({\rm Gi})^{1/2}%
}{\tau} \Biggr )
\end{equation}

The correction remains small in the entire region where nonlinear effects
are important $Gi < \tau < ({\rm Gi})^{1/2}$.

Note, that real superconductors are always inhomogeneous. The finite value
of the transition width leads to the appearance of an effective depairing
factor [11]. And the value of this depairing factor can be large enough in
the units of $T{\rm Gi}$. In such a case Maki-Thompson contribution to
conductivity is small compared to Aslamazov-Larkin contribution in the full
temperature region.

\section*{6. Conclusion}

We see, that nonlinear fluctuation effects in kinetics phenomena are much
stronger than in thermodynamics. If external depairing factor is absent,
then the nonlinear effects lead to the saturation of Maki-Thompson
contribution to conductivity in the temperature region $\tau \leq ({\rm Gi}%
)^{1/2}$. In this temperature region Aslamazov-Larkin contribution becomes
even stronger and grows as $\sigma ^{a}/\sigma _{0}\approx {\rm Gi}%
^{3/2}/\tau ^{2}$. In a superconductor with a large enough external
depairing factor $\Gamma =\tau _{s}^{-1}>T({\rm Gi})^{1/2}$ or short energy
relaxation time $\tau _{\varepsilon }^{-1}>T({\rm Gi})^{1/2}$ the
Maki-Thompson contribution saturates in temperature region $T\tau \leq \Gamma $
or $T\tau \leq \tau _{\varepsilon }^{-1}$. The nonlinear effects are not
very important for it. Magnetic impurities and energy relaxation act on
Aslamazov-Larkin contribution in a different way. Energy relaxation leads to
the appearance of collision integral in the kinetic equation for the
distribution functions of normal excitations. This collision integral
diminishes nonequilibrium contributions to the distribution functions.
Magnetic impurities and magnetic field act only on the superconductivity and
don't lead to the relaxation of the distribution functions. However, TDGL
equation depends essentially on the electron distribution function. If
parameter $\tau _{\varepsilon }^{-1}>T({\rm Gi})^{1/2}$, then nonlinear
fluctuation effects are not essential and Aslamazov-Larkin contribution
remains the same $\sigma ^{a}/\sigma _{0}={\rm Gi}/\tau $ in full
temperature region $\tau >{\rm Gi}$. If inequality $\tau _{\varepsilon
}^{-1}<T({\rm Gi})^{1/2}$ is fulfilled, then the law $\sigma ^{a}/\sigma
_{0}\simeq (Gi)^{3/2}/\tau ^{2}$ takes place in the temperature region $%
T\tau >\tau _{\varepsilon }^{-1}$. In the region $(T\tau _{\varepsilon
})^{-1}>\tau >{\rm Gi}$ correction to conductivity is given by the
expression $\sigma ^{a}/\sigma _{0}\sim {\rm Gi}^{3/2}T\tau _{\varepsilon
}/\tau $ (see Eq.(\ref{52})). Magnetic impurities (or current) suppress
nonlinear fluctuation effects in $\sigma ^{a}$, but the effect is not as
strong as in the case of energy relaxation. In the range $TGi/\Gamma >\tau >%
{\rm Gi}$ the correction to conductivity $\sigma ^{a}$ is given by Eq.(\ref
{59}) $\sigma ^{a}/\sigma _{0}\sim TGi^{2}/(\Gamma \tau ^{2})$. In the
temperature region $\tau >T{\rm Gi}/\Gamma $ the correction $\sigma ^{a}$
has the form (\ref{43}) in the linear approximation.

It is essential, that conductivity of fluctuating pairs can be larger then
conductivity of normal electrons in the temperature region, where correction
to the thermodynamic quantities is still small (see Eq.(61)).

A.I. Larkin thanks M.Yu. Reizer and V.M. Galitski for discussions. This work
(AL) was supported by NSF grant DMR-9812340.

The research of Yu.N. Ovchinnikov was made possible in part by Award No.
RP1-2251 of the U.S. Civilian Research $\&$ Development Foundation for the
Independent States of the Former Soviet Union (CRDF). Research of Yu.N.O.
supported also by RFBR.

\end{document}